# High-resolution structure of viruses from random diffraction snapshots

A. **Hosseinizadeh**,[1*] P. **Schwander**,[1*] A. **Dashti**,[2] R. **Fung**,[1] R.M. **D'Souza**,[2] and A. **Ourmazd**[1#]

[1]Dept. of Physics, University of Wisconsin Milwaukee, 1900 E. Kenwood Blvd, Milwaukee, WI 53211, USA

[2]Dept. of Mechanical Engineering, University of Wisconsin Milwaukee, 1900 E. Kenwood Blvd, Milwaukee, WI 53211, USA

* These authors contributed equally

*Corresponding author:* A. Ourmazd; Tel: (414) 229-2610; Email: Ourmazd@uwm.edu

*Subject terms:* Macromolecular assemblies, symmetry, X-ray lasers, manifold embedding, dimensionality reduction

Total words: 4,085

**Abstract**

The advent of the X-ray Free Electron Laser (XFEL) has made it possible to record diffraction snapshots of biological entities injected into the X-ray beam before the onset of radiation damage. Algorithmic means must then be used to determine the snapshot orientations and thence the three-dimensional structure of the object. Existing Bayesian approaches are limited in reconstruction resolution typically to 1/10 of the object diameter, with the computational expense increasing as the eighth power of the ratio of diameter to resolution. We present an approach capable of exploiting object symmetries to recover three-dimensional structure to high resolution, and thus reconstruct the structure of the satellite tobacco necrosis virus to atomic level. Our approach offers the highest reconstruction resolution for XFEL snapshots to date, and provides a potentially powerful alternative route for analysis of data from crystalline and nanocrystalline objects.

(139 words)

**1. Introduction**

Ultrashort pulses from X-ray Free Electron Lasers (XFEL's) have recently made it possible to record snapshots before the object is damaged by the intense pulse [1, 2]. This has, for example, resulted in de novo determination of protein structure from nanocrystals fabricated in vivo [3]. The ultimate goal, however, remains the determination of the three-dimensional (3D) structure of *individual* proteins and viruses [4], and their conformations [5]. This requires the ability to recover structure from an ensemble of ultralow-signal diffraction snapshots of unknown orientation. The 3D diffracted intensity can then be reconstructed, from which the real-space structure is recovered by iterative phasing algorithms [6-9].

The algorithmic challenge of determining XFEL snapshot orientations was first solved by iterative Bayesian approaches [10, 11], which assign an orientation to each snapshot based on maximum likelihood. A key measure of algorithmic performance is computational cost, which determines the range of amenable problems. Orientation recovery methods typically scales as $R^n$ per iteration, with the magnitude and scaling of the number of iterations unknown [12, 13]. For Bayesian algorithms, $n = 8$ [6, 7, 12-14], limiting the amenable resolution to ~1/10 of the object diameter [10-12]. At this level, biologically relevant study of almost all interesting objects such as proteins and viruses is out of reach. More recent methods offer improved performance, either by obviating the need for iteration [13], or by improved scaling per iteration, e.g., $\left(R^5 \log R\right)$ [14], though the magnitude and scaling of the number of iterations, where needed, remain unknown.

Despite these developments, computational expense remains a primary challenge. The highest resolution reported to date by methods conforming to the Shannon-Nyquist sampling theorem is ~1/30 of the object diameter [13, 15]. This is still inadequate for protein assemblies such as viruses. As viruses are expected to scatter more strongly than single molecules, they are under intense study by XFEL methods. High-resolution reconstruction of the 3D structure of a virus from XFEL snapshots thus represents an important milestone in the campaign toward single molecules. For all structure recovery techniques, the exploitation of symmetry offers an important and hitherto unused weapon in this endeavor.

The Shannon-Nyquist sampling theorem links the resolution $r$ with which an object can be reconstructed to the number of available snapshots $N_{snap}$, the object diameter $D$, and the number of elements $N_G$ in the point group of a symmetric object [12]:

$$r = \left( \frac{8\pi^2}{N_G N_{snap}} \right)^{1/3} D \quad . \tag{1}$$

This equation shows that the presence of symmetry can substantially increase the achievable resolution, or reduce the number of snapshots needed to achieve a certain resolution. The current experimental concentration on strongly scattering giant viruses [16] (large $D$), and the scarcity of "useful" single-particle snapshots [17] (small $N_{snap}$) make the exploitation of symmetry crucial for further progress. No reconstruction method capable of operating at the signal-to-noise ratios expected from single-molecule diffraction has, to date, exploited object symmetry.

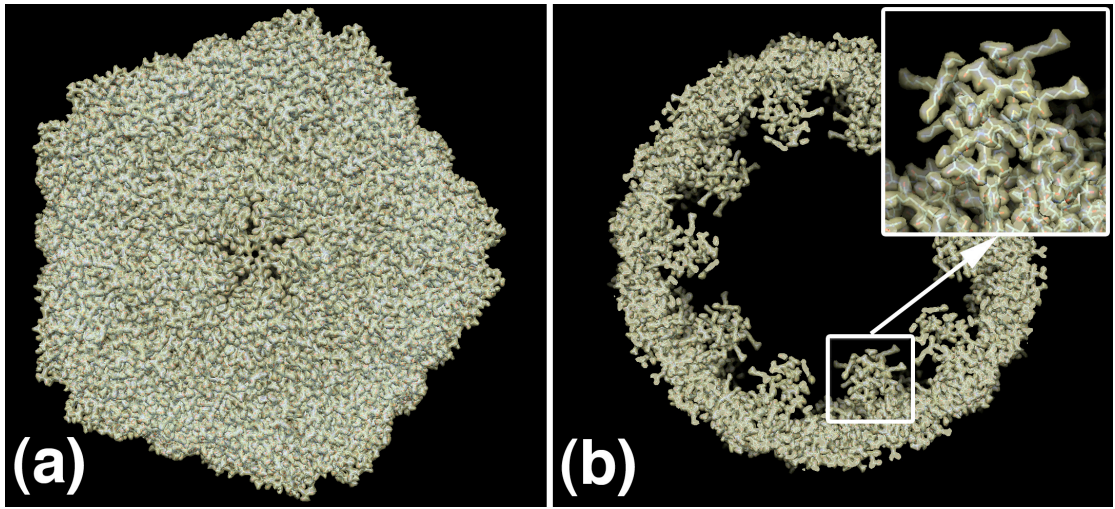

**Figure 1.** Recovered structure of the satellite tobacco necrosis virus to atomic resolution. The signal is one scattered photon per Shannon pixel at 0.2nm, with Poisson (shot) noise included. This corresponds to the signal expected from viruses currently under investigation with XFEL techniques. (a) The three-dimensional electron density extracted from 1.32 million noisy diffraction snapshots of unknown orientation, demonstrating structure recovery to a resolution 1/100 of the object diameter (here 0.2nm). (b) A slice of the electron density showing atomic resolution. The known structure is shown as a ball-and-stick model without adjustment (inset).

Here, we present an approach capable of determining structure from diffraction snapshots of symmetric objects to 1/100 of the object diameter, and demonstrate 3D structure recovery to atomic resolution from simulated noisy snapshots of the satellite tobacco necrosis virus at the signal level expected from viruses currently under study at the LCLS X-ray Free Electron Laser (Fig. 1). This approach can be applied to symmetric objects of any kind, opening the way to the high-resolution study of a wide variety of crystalline and non-crystalline biological and non-biological entities without radiation damage.

Due to the superior computational efficiency and hence reconstruction capability of non-iterative manifold approaches [13], we focus on incorporating this capability into these powerful algorithms [13, 15, 18, 19]. In brief, these approaches recognize that scattering

"maps" a given object orientation to a diffraction snapshot. The collection of all possible orientations in 3D space spans an SO(3) manifold. Scattering maps this manifold to a topologically equivalent compact manifold in the space spanned by the snapshots. We have shown that, to a good approximation, the manifold formed by the snapshots is endowed with the same metric as that of a "symmetric top", loosely speaking a sphere squashed in the direction of the incident beam due to the effect of projection [13]. Such a manifold is naturally described by the Wigner *D*-functions [20], which are intimately related to the elements of the (3x3) rotation matrix [13]. Via so-called Empirical Orthogonal Functions, powerful graph-based algorithms [18, 21, 22] provide access to the Wigner *D*-functions describing manifolds produced by scattering [13], from which the snapshot orientations can be extracted [13, 15].

It is the object of this paper to incorporate object symmetry into manifold-based approaches, and thus enable high-resolution structure recovery by XFEL methods. We show that the Diffusion Map algorithm [21], a theoretically sound (in the sense of guaranteed convergence to eigenfunctions of known operators) and algorithmically powerful (in the sense of intrinsic sparsity) manifold-based approach can be used to recover structure from random snapshots of a symmetric object to high resolution. For concreteness, the discussion is restricted to icosahedral objects, but the approach can be applied to any crystalline or non-crystalline object with symmetry.

The paper is organized as follows. Section 2 outlines our theoretical approach. Specifically, it addresses the construction of eigenfunctions suitable for manifolds produced by scattering from symmetric objects, and describes how symmetry-related ambiguities in orientation recovery may be resolved. Section 3 demonstrates structure recovery from simulated diffraction snapshots of a symmetric object to 1/100 of its diameter. For the satellite tobacco necrosis virus used as example, this corresponds to atomic resolution. Section 4 places our work in the context of ongoing efforts to determine structure by scattering from single particles and nanocrystals. Section 5 concludes the paper with a brief summary of the implications of our work for structure determination by XFEL techniques. Theoretical and algorithmic details, including pseudocode are presented as Electronic Supplementary Material (ESM).

**2. Theoretical approach**

We begin by constructing the eigenfunctions needed to describe manifolds produced by scattering from symmetric objects. Diffusion Map describes a manifold in terms of the eigenfunctions of the Laplace-Beltrami operator with respect to an unknown metric [13, 18]. In the absence of object symmetry, manifolds produced by scattering are well approximated by a homogeneous metric, with the eigenfunctions of the Laplace-Beltrami operator corresponding to the Wigner $D$-functions [13, 15]. In the presence of object symmetry, appropriately symmetrized eigenfunctions are needed. As shown in ESM sections A and B, these can be obtained by summing over the Wigner $D$-functions after operation by the elements of the object point-group, viz.

$$\hat{D}^j_{m',m}(\alpha) = \frac{1}{N_G} \sum_{R_i \in G} O_{R_i} D^j_{m',m}(\alpha) \quad , \tag{2}$$

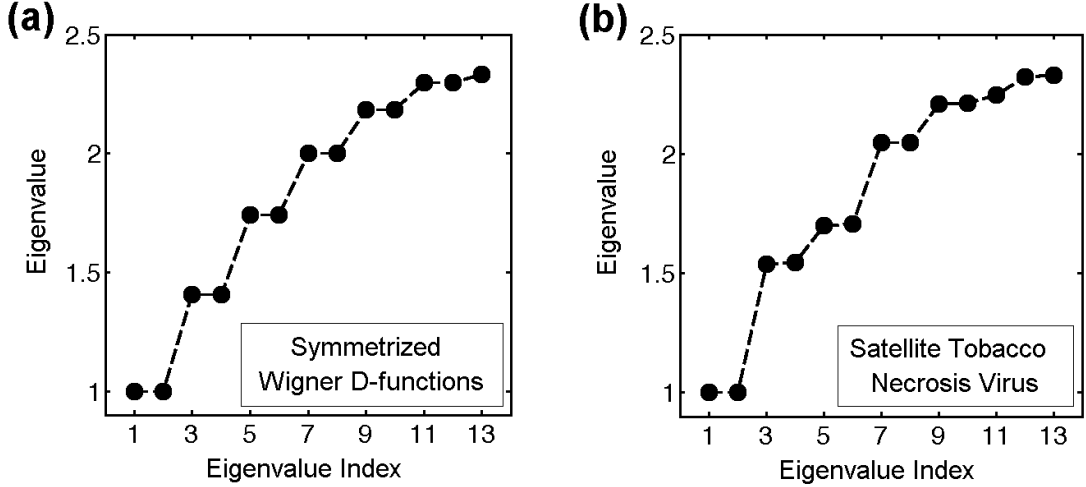

**Figure 2.** Eigenvalue spectra of the Laplace-Beltrami operator. (a) Spectrum for icosahedral Wigner *D*-functions. (b) Spectrum obtained from the manifold produced by noise-free simulated diffraction snapshots of the satellite tobacco necrosis virus. Note the close agreement between the two spectra.

where $\alpha$ denotes the three numbers collectively representing any rotation, $N_G$ the number of operations $O_{R_i}$ in the point-group $G$, and $D^j_{m',m}(\alpha)$ the (real) Wigner *D*-functions. This approach is applicable to all point groups. For the icosahedral group, the lowest allowed eigenfunctions consist of 39 non-zero $\hat{D}^6_{m',m}(\alpha)$, whose orthogonalization leads to 13 independent icosahedral functions $\tilde{D}^6_m(\alpha), (-6 \leq m \leq 6)$ (see ESM section B.) These comprise one non-degenerate ($m = 0$) and six degenerate pairs of eigenfunctions, with the *m* in each pair differing only in sign (Fig. 2). A similar set of eigenvalues results from Diffusion Map analysis of diffraction snapshots. (The differences between the two sets of eigenvalues are most likely due to the homogeneous metric approximation [13].)

Direct comparison of the eigenvalues of icosahedral Wigner *D*-functions with those obtained from Diffusion Map analysis (designated here by $\psi_i$) is not a sufficiently reliable means of identifying each $\psi_i$ with its correct partner among the Wigner $D$-functions. This can be achieved by reference to plots of all snapshot coordinates for different pairs of $\psi_i$. These display characteristic patterns, from which each of the 13 $\psi_i$ can be reliably associated with one of the symmetrized Wigner $D$-functions $\tilde{D}_m^6(\alpha)$ (Fig. 3). (The plots corresponding to $m = \pm 3$ and $\pm 5$ are similar. However, snapshots with 5-fold symmetry occur at the center of the $m = \pm 3$ plot and along a circle in $m = \pm 5$, allowing unambiguous distinction.)

Next, we describe how orientations can be extracted from analysis of diffraction snapshots. In principle, once each of the first thirteen $\psi_i$ has been identified with a symmetrized eigenfunction $\tilde{D}_m^6(\alpha)$, the orientation of each snapshot can be extracted from its coordinates in the space spanned by the thirteen $\psi_i$. This is complicated, however, by the presence of symmetry, which introduces degeneracies in the symmetrized eigenfunctions, as outlined above. Clearly, all orthogonal and normalized degenerate $(\psi_i, \psi_j)$ pairs are equally acceptable. More precisely, any orthogonal operation on such a pair of eigenfunctions leads to another equivalent pair. Each degenerate pair $(\psi_i, \psi_j)$ is thus related to its counterpart $(\tilde{D}_m^6(\alpha), \tilde{D}_{-m}^6(\alpha))$ via an unknown mixing angle $\theta_m$, and a scaling factor, viz.

$$\begin{pmatrix} \tilde{D}_m^6 \\ \tilde{D}_{-m}^6 \end{pmatrix} = \frac{1}{\sqrt{13}} \begin{pmatrix} \cos\theta_m & (-1)^{m+1}\sin\theta_m \\ (-1)^m \sin\theta_m & \cos\theta_m \end{pmatrix} \begin{pmatrix} \psi_i \\ \psi_j \end{pmatrix} . \qquad (3)$$

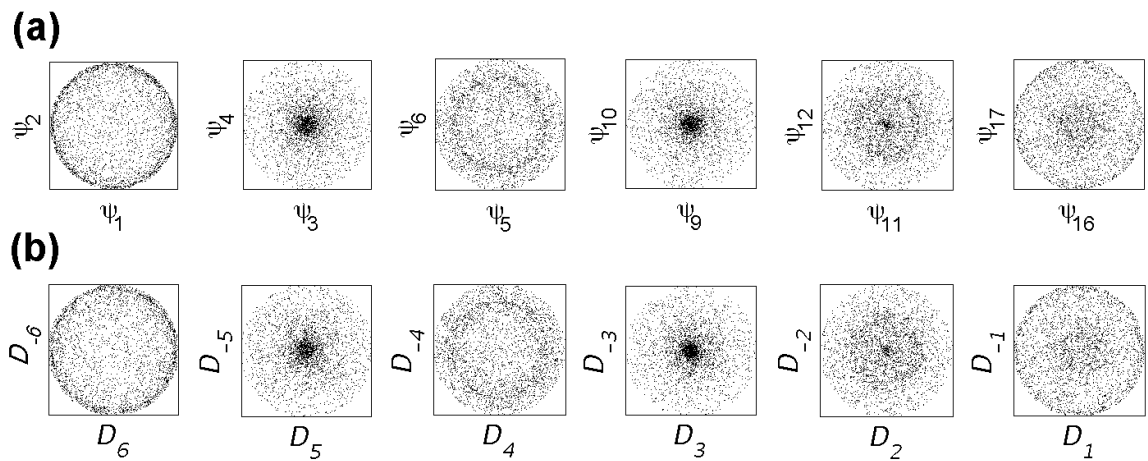

**Figure 3.** Scatter plots to identify eigenfunctions. These, together with the distribution of 5-fold symmetric snapshots allow unambiguous association of the Diffusion Map eigenvectors $\psi_i$ with their counterparts among the icosahedral Wigner $D$-functions $\tilde{D}_m^6$. (a) $\psi_i$ vs. $\psi_j$ plots of snapshot coordinates obtained from Diffusion Map. (b) $\tilde{D}_m^6$ vs. $\tilde{D}_{-m}^6$ plots for randomly sampled points in the space of orientations.

Additionally, it must be established whether an inversion operation should be inserted on the right side of Eq. (3).

The mixing angle for each of the six degenerate pairs can be thought of as the position of the hour hand on a clock. Ideally, one would like all six clocks to display Greenwich Mean Time (have the same mixing angle $\theta_m$). However, the arbitrary orthogonal operations allowed by the presence of degeneracy mean that each clock could show a different "local" time. As orthogonal operations also include inversion, the sense of rotation of each clock could also be reversed. One must therefore determine the local time and the sense of clock rotation in order to relate the Diffusion Map eigenfunctions $\psi_i$ to the symmetrized eigenfunctions $\tilde{D}_m^6(\alpha)$. As described in more detail in ESM section C, this can be accomplished as follows.

First, we describe how the orthogonal transformation of degenerate pairs, or, equivalently, the mixing angle $\theta_m$ can be determined. It can be easily shown that the rotation of an object through $\pi$ about the y-axis changes the position of a snapshot in a plot of real Wigner D-functions by a mirror operation about the line $\theta_m = 0$, viz.: $(\tilde{D}_m^6, \tilde{D}_{-m}^6) \rightarrow (\tilde{D}_m^6, -\tilde{D}_{-m}^6)$. The line corresponding to the zero of the mixing angle is, therefore, the perpendicular bisector of the line connecting the coordinates of a given snapshot and that produced by rotating the object by $\pi$ about the y-axis (snapshots $a$ and $\bar{a}$ in Fig. 4). As shown in Fig. 5, in the presence of Friedel symmetry, the snapshot $\bar{a}$ can be simply produced by mirroring $a$ about the detector x-axis. The mixing angle can thus be determined to within $\pi$ by adding the appropriate mirror images of a subset of the snapshots to the dataset before Diffusion Map embedding (Fig. 4). The remaining $\pi$ ambiguity stems from the sense chosen for the perpendicular bisector, and is resolved later (see below).

Next, we describe how the presence of possible inversions (reversal of the sense of clock rotation) can be determined. Consider a subset of snapshots, and rotate each by a small amount about a central axis perpendicular to its plane to form a new subset of snapshots. Embed the augmented dataset. The sense of rotation can now be determined by observing whether a rotated snapshot leads or trails its unrotated counterpart.

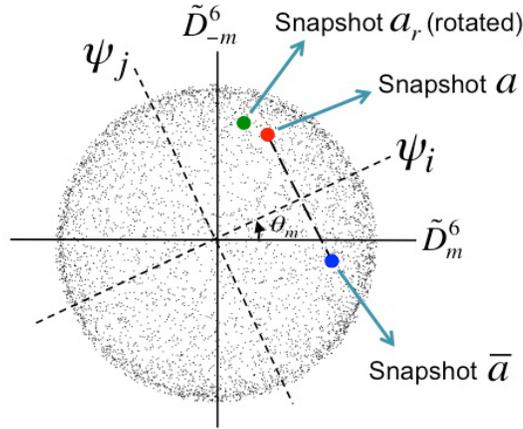

**Figure 4.** Schematic diagram showing the effect of the mixing angle $\theta_m$ between a pair of normalized degenerate eigenfunctions. Each point represents the coordinates of a snapshot. The zero of the mixing angle is given by the perpendicular bisector of the line connecting snapshots $a$ and $\bar{a}$ as described in the text. The sense of clock rotation can be determined from the position of a snapshot rotated by a few degrees about the beam axis.

A link is now established between the Diffusion Map eigenfunctions $\psi_i$ and the symmetrized Wigner $D$-functions $\tilde{D}_m^6(\alpha)$, to within a $\pi$ ambiguity in each of the six mixing angles. The snapshot orientations can now be extracted from the $\psi_i$ by a least-squares fit in a straightforward manner, as described in detail in ESM section D. The $\pi$ ambiguity is resolved by performing fits for each of the 64 ($2^6$) possibilities, and selecting the outcome with the lowest residual.

## 3. Results

We now demonstrate our approach by reconstructing the structure of an icosahedral virus to 1/100 of its diameter both with and without noise. For the satellite tobacco necrosis virus (STNV, PDB designation: 2BUK) used here, this corresponds to atomic resolution (0.2 nm).

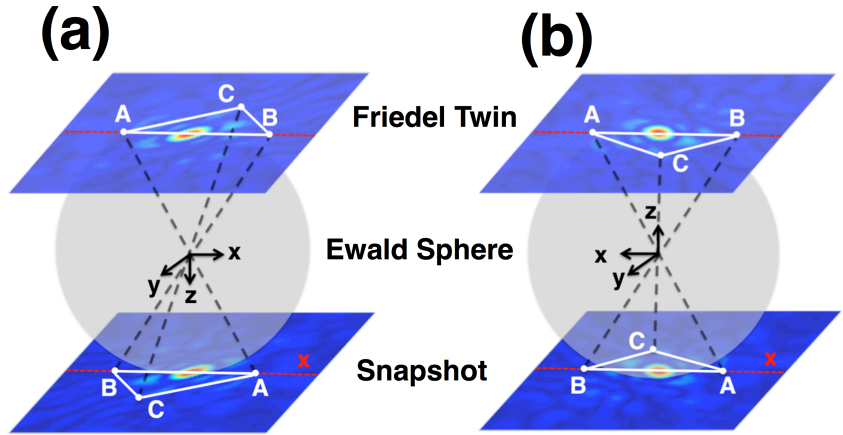

**Figure 5.** Schematic diagram showing a snapshot and its Friedel twin (a) before, and (b) after an object rotation through $\pi$ about the y-axis. The snapshots at the bottom are related by a mirror operation about the detector x-axis.

To estimate the signal expected from viruses, we calculated the number of elastically scattered photons per Shannon pixel for the STNV [23], one of the smallest viruses, and for the paramecium bursaria chlorella virus [24], one of the larger viruses known. (See Table, and ESM section E). The exact number of scattered photons depends on a number of parameters, but for each virus, it can be estimated from the number and energy of incident photons, the beam diameter, and the maximum scattering angle. At photon energies typically used for XFEL studies of viruses, the number of photons scattered to a Shannon pixel at 30° collection angle ranges from ~1 to 3000. The signal level producing one scattered photon per Shannon pixel at 30° was therefore used to simulate Poisson noise. The resulting signal-to-noise ratio is well above those amenable to our approach without a denoising step [14].

**Table**

Number of photons scattered to a Shannon pixel at 30° by a small and a large virus
The beam diameter is matched to the object size (STNV: 20nm; Chlorella: 190nm)

| Virus | Photon Energy (keV) | Photons/Pulse ($mm^{-2}$) | (Obj. Diameter / Resolution) | *Scattered Photons/Shannon Pixel* |
|---|---|---|---|---|
| STNV | 0.5 | $10^{13}$ | 4 | **3770** |
| | 2 | $4 \times 10^{12}$ | 17 | **89** |
| | 5 | $10^{12}$ | 42 | **3** |
| | 7 | $10^{12}$ | 58 | **1** |
| Chlorella | 0.5 | $10^{13}$ | 40 | **280** |
| | 2 | $4 \times 10^{12}$ | 158 | **7** |
| | 2.5 | $10^{12}$ | 198 | **1** |

We now demonstrate the performance of our approach with reference to simulated snapshots of STNV to 0.2 nm (crystallographic) resolution (corresponding to 1/100 of the object diameter) at an incident photon energy of 12.4 keV. The demonstration includes two sets of simulated data: one noise-free; the second including shot-noise corresponding to a mean of one photon per Shannon pixel at 0.2nm. (For details, see ESM section E.) The conditions are chosen to highlight the noise robustness and resolution of our approach. (The Chlorella virus at 2.5 keV would have served equally well.) The appropriate experimental conditions, of course, depend on a number of additional parameters, such as differential scattering with respect to the solvent, etc.

**4. Discussion**

We now outline the primary implications of our work. Incorporation of object symmetry has proved a powerful tool for recovering structure by established single-particle techniques, such as cryo electron microscopy (cryo-EM) [25]. By enhancing the effective number of snapshots and improving resolution, our approach promises to play a similarly important role in 3D structure recovery by XFEL methods. In terms of resolution expressed as a fraction of the object diameter, our approach is comparable with the best achieved by cryo-EM approaches [26], but without phase information. Combined with its superior noise robustness [15, 17], our approach offers a vital route to determining high-resolution structure at signal levels expected even from single macromolecules in XFEL experiments [10, 15, 27]. For biological entities in particular, this is essential for obtaining "biologically relevant" information.

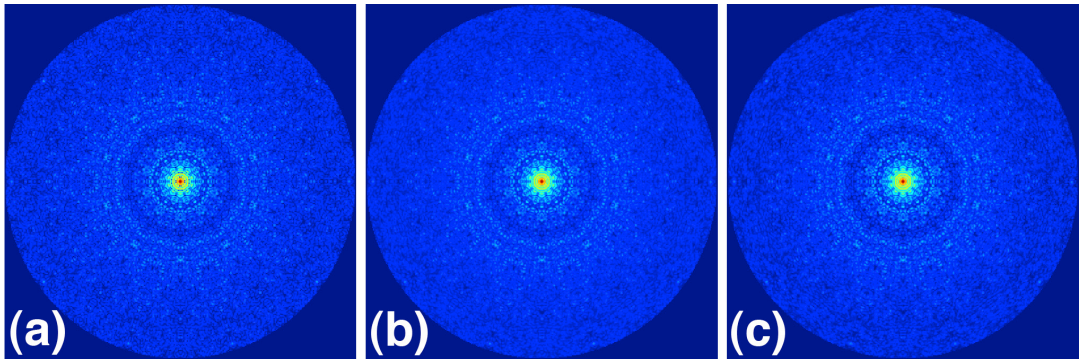

**Figure 6.** Comparison of the exact and recovered diffraction volumes. (a) A slice through the exact 3D diffraction volume. Same slice through the recovered diffraction volumes (b) without noise, and (c) with Poisson noise corresponding to a mean signal of one photon per Shannon pixel at a resolution corresponding to 1/100 of the object diameter.

XFEL experiments to obtain snapshots from individual biological objects are in progress [16, 28]. The only publicly available XFEL dataset on viruses [28], however, suffers from the presence of experimental artifacts, such as variations in beam intensity, position and inclination, and limitations due to detector dynamic range and nonlinearities. The rapid progress in XFEL-based nanocrystallography [2] leads us to expect improved single-particle datasets quickly. By exploiting object symmetry, our manifold-based approach thus represents a vital and timely tool for high-resolution structure recovery from symmetric, biological and non-biological single particles by XFEL methods.

More generally, our approach can be applied also to structure recovery by XFEL-based nanocrystallographic methods. Traditional indexing approaches, combined with Monte Carlo integration techniques have provided impressive first results [1-3]. However, issues such as the so-called "twinning ambiguity" and the effect of variations in nanocrystal size and shape have so far eluded resolution. The incorporation of symmetry into manifold-based orientation recovery offers the possibility to avoid this ambiguity by obviating the need for index-based orientation of crystalline diffraction patterns.

## 5. Summary and conclusions

We have demonstrated the first approach capable of extracting high-resolution 3D structure from diffraction snapshots of symmetric objects, and presented structure recovery to 1/100 of the object diameter at signal-to-noise ratios expected from currents XFELs. This opens the way to the study of individual biological entities before the onset of significant radiation damage. Our approach also offers the possibility to apply powerful graph-theoretic techniques to the study of crystalline objects, with the potential to extract more information from the rich and rapidly growing body of nanocrystallographic data.

**Acknowledgments**

We are grateful to Dimitrios Giannakis and Dilano Saldin for valuable discussions. This research was supported by: the US Dept. of Energy, Office of Science, Basic Energy Sciences under award DE-FG02-09ER16114 (overall design); the US National Science Foundation under awards MCB-1240590 (algorithm development), CCF-1013278 and CNS-0968519 (GPU algorithms); and the UWM Research Growth Initiative (theory). The publication of this work was supported by the US National Science Foundation under award STC 1231306.

free-electron laser x-ray diffraction data sets for algorithm development. *Optics Express* **20**(4), 4149-4158. (doi: 10.1364/OE.20.004149).